\documentclass{PoS}

\usepackage{color}
\usepackage{refstyle}
\usepackage{amsmath}
\usepackage{amssymb}
\usepackage{graphicx}
\usepackage{wasysym}
\usepackage{cite}

\AtBeginDocument{}
\AtBeginDocument{\providecommand\eqref[1]{\ref{eq:#1}}}
\AtBeginDocument{}
\AtBeginDocument{}

\makeatother

\title{Combining ordinary and topological finite volume effects for fixed 
topology simulations}

\ShortTitle{ Ordinary and Topological finite volume effects}

\author{\speaker{Arthur Dromard}$^{(1)}$, Wolfgang Bietenholz$^{(2)}$, Urs 
Gerber$^{(2,3)}$, \newline H\'{e}ctor Mej\'{\i}a-D\'{\i}az$^{(2)}$ and 
Marc Wagner$^{(1)}$ \\
\vspace*{1mm}
\ \\
$^{(1)}$~Goethe-Universit\"at Frankfurt am Main \\
~~~~ Institut f\"ur Theoretische Physik \\
~~~~ Max-von-Laue-Stra{\ss}e 1, D-60438 Frankfurt am Main, Germany
\vspace*{1.5mm} \\
$^{(2)}$~Instituto de Ciencias Nucleares \\
~~~~ Universidad Nacional Aut\'{o}noma de M\'{e}xico \\
~~~~ A.P.\ 70-543, C.P.\ 04510 Distrito Federal, Mexico
\vspace*{1.5mm} \\
$^{(3)}$ Instituto de F\'{\i}sica y Matem\'{a}ticas\\
~~~~ Universidad Michoacana de San Nicol\'{a}s de Hidalgo\\
~~~~ Edificio C-3, Apdo.\ Postal 2-82, C.P.\ 58040, 
Morelia, Michoac\'{a}n, Mexico \vspace*{3mm} \\
        E-mail: 
\email{dromard@th.physik.uni-frankfurt.de}}
\abstract{In lattice quantum field theories with topological sectors,
simulations at fine lattice spacings --- with typical algorithms ---
tend to freeze topologically. In such cases, specific topological finite size 
effects have to be taken into account to obtain physical results, which
correspond to infinite volume or unfixed topology. 
Moreover, when a theory like QCD is
simulated in a moderate volume, one also has to overcome ordinary 
finite volume effects (not related to topology freezing). To extract physical 
results from simulations affected by both types of finite volume effects, we 
extend a known relation between hadron masses at fixed and 
unfixed topology by additionally incorporating ordinary finite volume 
effects. We present numerical results for SU(2) Yang-Mills theory.}

\FullConference{The 33rd International Symposium on Lattice Field Theory\\
		14 -18 July 2015\\
		Kobe International Conference Center, Kobe, Japan*}

\begin{document}

\vspace{-1mm}
\section{Introduction }
\vspace{-1mm}

In order to extrapolate simulation results in lattice QCD
to the continuum limit, the lattice spacing is made as small
as possible. However, for an algorithm like Hybrid Monte Carlo, 
this implies an increase of the topological auto-correlation time, 
and finally a freezing of the topological charge, for any lattice 
discretisation \cite{Luscher:2011kk}.
In the case of chirally symmetric fermions, such as overlap fermions,
the Monte Carlo history tends to get stuck in one topological sector
even for rather large lattice spacing, see {\it e.g.}\ 
Refs.\ \cite{Aoki:2008tq,Borsanyi}. In
specific cases it might be motivated to fix topology on purpose.
For example when using a mixed action setup with overlap valence
and Wilson sea quarks, one observes an ill-behaved continuum limit
due to different near zero-modes in the sea and valence quark 
sector\cite{Cichy:2010ta}. Extracting
physically meaningful results from the topologically trivial sector
could be a solution. 

There are several strategies to deal with this problem. 
The method we are interested in is based on a relation between 
the physical mass of a hadron and its fixed topology counterpart 
\cite{Brower:2003yx,Aoki:2007ka,Dromard:2014ela}. This method has already
been tested with success in various 
models\cite{Bietenholz:2011ey,Bietenholz:2008rj,Bietenholz:2012sh,
Dromard:2013wja,Czaban:2013haa,Bautista:2014tba,
Czaban:2014gva,Dromard:2014gma,Gerber:2014bia,Bautista:2015yza}. 
So far ordinary finite volume effects have been neglected. 
However, also these effects can be important for QCD simulations with 
light pions. Here we describe an extension of 
the method of Ref.\ \cite{Brower:2003yx},
combining both topological and ordinary finite volume 
effects for SU($N$) Yang-Mills theories, as well as QCD, 
as anticipated in Ref.\ \cite{Dromard:2015oqa}.
After a brief discussion of ordinary finite volume effects and the method used
to extract masses from fixed topology, we present the extended equations
and the results of a numerical test in SU(2) Yang-Mills theory. 

\vspace{-1mm}
\section{Ordinary finite volume effects\label{Ordinary-finite-volume}}
\vspace{-1mm}

The difficulties in lattice simulations
related to finite volume effects are well-known. 
A finite volume with periodic boundary conditions allows
the particle to interact with its own copies.
This artificial interaction causes a shift in the value of the 
mass obtained by simulations. 

This mass shift has been rigorously analysed in Ref.\ \cite{Luscher:1985dn}. 
Different results were obtained for SU($N$) gauge theory and for QCD,
due to the different masses of the lightest stable particle, namely 
the $0^{++}$ glueball and the pion. %, respectively.
For Yang-Mills theories one obtains % the relation
%%%%%%%%%%%%%%%%%%%%%%%%%%%%%%%%%%%%%%%%%%%%%%%%%%%%%%
\begin{equation}
M(L)-M(L=\infty) \propto \frac{1}{L}
\exp \Bigg( -\frac{\sqrt{3}m_{\rm g}L}{2} \Bigg) \ ,
\label{eq:SU(N)}
\end{equation}
%%%%%%%%%%%%%%%%%%%%%%%%%%%%%%%%%%%%%%%%%%%%%%%%%%%%%%
where $m_{\rm g}$ is the mass of the lightest glueball, %{\it i.e.}\ 
($J^{PC}=0^{++}$) and $L$ is the spatial extent of the lattice. The corresponding QCD relation reads
%%%%%%%%%%%%%%%%%%%%%%%%%%%%%%%%%%%%%%%%%%%%%%%%%%%%%%
\begin{equation}
m_{\pi}(L)-m_{\pi}(L=\infty)\propto\frac{1}{L} 
K_{1}(m_{\pi}L)\label{eq:QCD} \ ,
\end{equation}
%%%%%%%%%%%%%%%%%%%%%%%%%%%%%%%%%%%%%%%%%%%%%%%%%%%%%%
where $m_{\pi}$ is the pion mass and $K_{1}$ the modified Bessel function.
We confronted eq.\ (\ref{eq:SU(N)}) with lattice SU(2) Yang-Mills
results for the static quark-antiquark potential
$\hat{\mathcal{V}}_{q\bar{q}}$ (hats denote quantities
in lattice units). For this purpose, we
computed $\hat{\mathcal{V}}_{q\bar{q}}(\hat r = 3)$ in the volumes
$\hat V = \hat L^{4}$, $\hat L = 11,12, \dots , 16$ and $18$, 
based on $4000$ configurations in each volume.\footnote{The same
set of configurations was used for the results reported in Section 3
to 5. Identifying the Sommer  parameter $r_{0}$ with $0.46~{\rm fm}$ one obtains a lattice spacing
of $a \approx 0.073~{\rm fm}$.} The results
are in excellent agreement with the theory. For $\hat{L}\geq14$
the finite volume effects are negligible,
but in smaller volumes they have to be taken into account.
From the fit of eq.\ (\ref{eq:SU(N)})
to the lattice results we can extract the glueball 
mass to a good precision, $\hat{m}_{\rm g}=0.74(4)$ (cf. Figure 2 in \cite{Dromard:2015oqa})
This is in excellent agreement with the result obtained by a standard
lattice computation of a glueball 2-point correlation function, 
$\hat{m}_{\rm g} = 0.723(23)$ \cite{Teper:1998kw}.

\vspace{-1mm}
\section{Topological finite volume effects\label{sec:Topological-finite-volume}}
\vspace{-1mm}

Fixing the topology, {\it i.e.}\ the restriction to only one topological 
sector, entails additional typically even stronger finite volume effects. A relation
between a physical hadron mass $M$ and the hadron mass
$M_{Q,V}$ obtained at fixed topological charge $Q$, in a finite
space-time volume $V$, has been derived in Ref.\ \cite{Brower:2003yx},
%%%%%%%%%%%%%%%%%%%%%%%%%%%%%%%%%%%%%%%%%%%%%%%%%%%%%%
\begin{equation}
M_{Q,V}=M+\frac{1}{2\chi_{t}V}M^{(2)}\Bigg(1-\frac{Q^{2}}{\chi_{t}V}
\Bigg)+\mathcal{O}\Bigg(\frac{1}{\left(\chi_{t}V\right)^{2}}\Bigg),
\label{eq:BCNW}
\end{equation}
%%%%%%%%%%%%%%%%%%%%%%%%%%%%%%%%%%%%%%%%%%%%%%%%%%%%%%
where $M^{(2)}$ is the second derivative of $M$ with respect to
the vacuum angle $\theta$, at $\theta=0$, and 
$\chi_{t}$ is the topological susceptibility.
One can see that fixed topology causes finite volume effects,
which are only suppressed by powers of the inverse volume. 

\begin{figure}
\begin{centering}
\includegraphics[width=10cm]{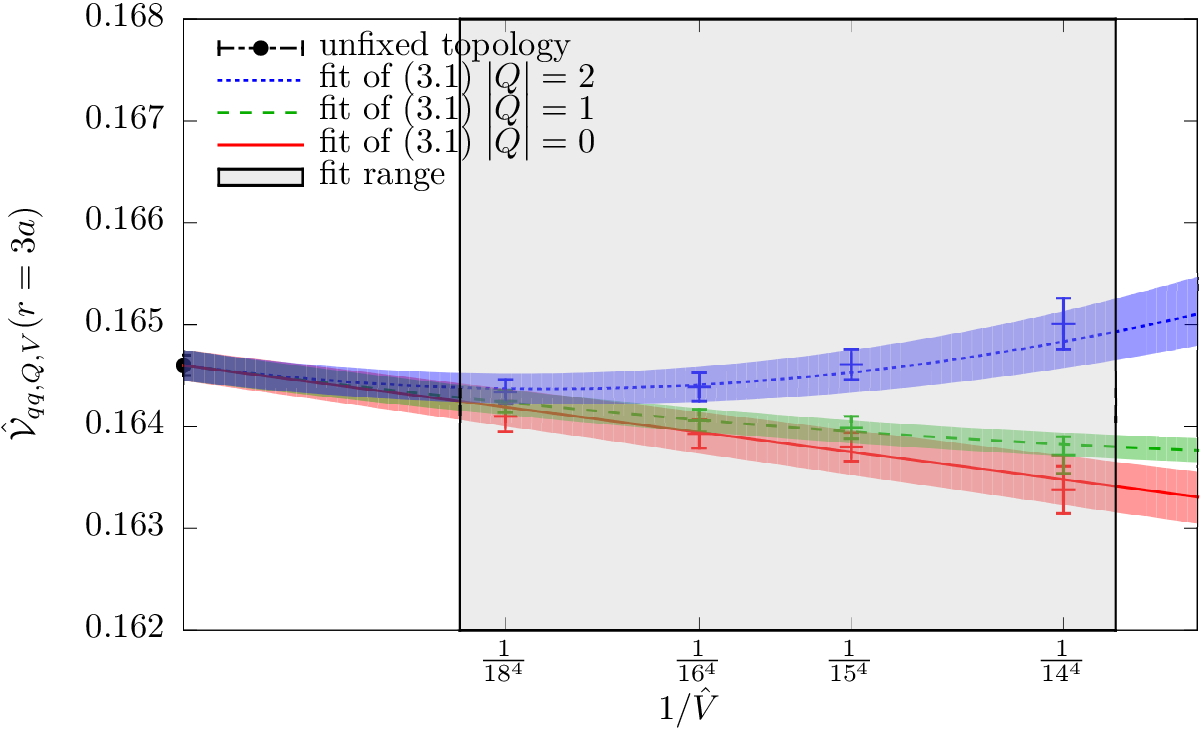}
\par\end{centering}
\protect\caption{\label{figR=3}The static potential (in lattice units),
$\hat{\mathcal{V}}_{q \bar{q},Q,V}(\hat r = 3)$, as a function of the inverse
lattice volume, $1/\hat{V}$. The curves 
represent the fit of eq.\ (\protect\ref{eq:BCNW}) to the lattice results 
in the volumes $\hat{V} = 14^4, 15^4, 16^4, 18^4$.}
\vspace{-2mm}
\end{figure}

In practice, the physical mass $M$ can be extracted
by a fit of eq.\ (\ref{eq:BCNW}) to numerical results for $M_{Q,V}$
in various volumes and topological sectors.
This method has been tested in Refs.\ 
\cite{Bietenholz:2011ey,Bietenholz:2008rj,Bietenholz:2012sh,
Dromard:2013wja,Czaban:2013haa,Bautista:2014tba,Dromard:2014ela,
Czaban:2014gva,Dromard:2014gma,Gerber:2014bia,Bautista:2015yza,
Dromard:2015oqa} with success. 
% Here we summarise recent results for SU(2) Yang-Mills theory,
% which have been presented previously in Ref.\ \cite{Dromard:2015oqa}. 
The static potential $\hat{\mathcal{V}}_{q\bar{q},Q,V}$
has been computed for different separations $(\hat r=1\ldots6)$ in 
different topological sectors, $|Q|=0,1,2$, and various volumes,
$\hat V=14^{4},15^{4},16^{4},18^{4}$.
Figure \ref{figR=3} shows $\hat{\mathcal{V}}_{q\bar{q},Q,V}(\hat r = 3)$;
there is a clear distinction between topological sectors, 
especially in small volumes. This calls for an indirect 
method to extract physical results.
The curves %in Figure \ref{figR=3} 
represent a global fit of eq.\ (\ref{eq:BCNW}) to the numerical results,
which is of good quality, $\chi^{2}/{\rm d.o.f.} \lesssim 1$.
The extrapolated potential $\hat{\mathcal{V}}_{q\bar{q}}(\hat r=3)=0.1646(2)$ 
is in excellent agreement with the corresponding straight 
computation at unfixed topology,
$\hat{\mathcal{V}}_{q\bar{q}}(\hat r=3)=0.16455(7)$.
%%%%%%%%%%%%%%%%%%%%%%%%%%%%%%%%%%%%%%%%%%%%%%%%%%%%%%
%%%%%%%%%%%%%%%%%%%%%%%%%%%%%%%%%%%%%%%%%%%%%%%%%%%%%%

In $\sigma$-model studies, also $\hat \chi_{t}$ could be evaluated well
in this manner \cite{Bautista:2014tba,Gerber:2014bia}.
In 4d Yang-Mills theory, however, the determination of $\hat \chi_{t}$ 
turned out to be plagued by large statistical errors.
% which implies that the method is not suitable to compute 
% the value of this latter.
A more promising method for that purpose has been suggested in Ref.\ 
\cite{Aoki:2007ka}, which showed that the correlation of the
topological charge density $q(x)$ at fixed topology behaves as
%%%%%%%%%%%%%%%%%%%%%%%%%%%%%%%%%%%%%%%%%%%%%%%%%%%%%%
\begin{equation}
\left\langle q(x)q(0)\right\rangle_{Q,V} \underset{|x|\rightarrow\infty}{\approx}
-\frac{\chi_{t}}{V} \Big( 1-\frac{Q^{2}}{\chi_{t}V} \Big)
+ \mathcal{O} \Big(\frac {1}{V^{2}} \Big) \ .
\label{eq:AFHO}
\end{equation}
%%%%%%%%%%%%%%%%%%%%%%%%%%%%%%%%%%%%%%%%%%%%%%%%%%%%%%
Eq.\ (\ref{eq:AFHO}) implies that --- at a large separation $|x|$ ---
the correlation $\left\langle q(x)q(0)\right\rangle _{Q,V}$ should
reach a plateau, and that its value determines $\chi_{t}$. The advantage
of this method is the need of only one volume and one topological sector. 

\begin{figure}
\vspace*{-4mm}
\begin{centering}
\includegraphics[angle=-90,width=10cm]{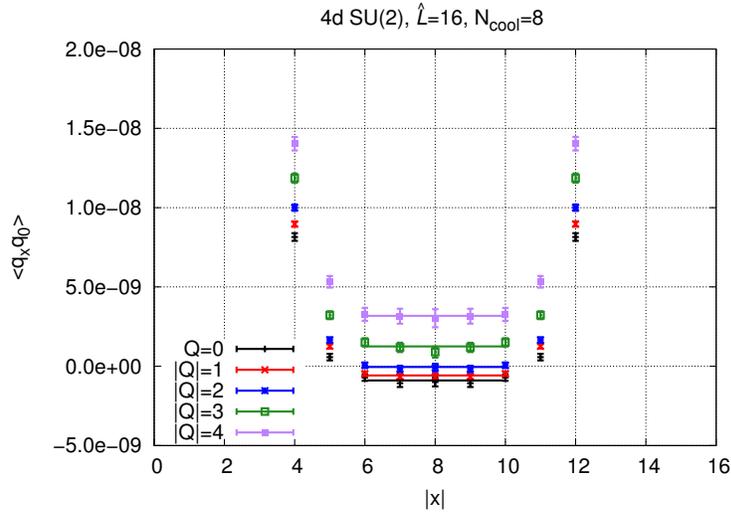}
\par\end{centering}
\protect\caption{\label{fig:Topological-density-correlator}The correlation 
$\langle \, q_x \, q_0 \, \rangle_{|Q|}$ 
as a function of the on-axis separation $|x|$, after 
$8$ cooling sweeps, for the lattice volume $\hat V = 16^4$.}
\vspace{-6mm}
\end{figure}

Figure \ref{fig:Topological-density-correlator} has been generated
%using the aforementioned configurations 
in the volume $\hat V=16^{4}$, based on
an all-to-all computation of the correlator. This was carried out 
after performing 8 cooling sweeps, in order to smooth out the ultra-violet
fluctuations without destroying the topological structure. 
Each fit of the plateau works with $\chi^{2}/{\rm d.o.f.} \approx 1$. 
This suggests that this method is promising indeed.
%%%%%%%%%%%%%%%%%%%%%%%%%%%%%%%%%%%%%%%%%%%%%%%%%%%%%%
%%%%%%%%%%%%%%%%%%%%%%%%%%%%%%%%%%%%%%%%%%%%%%%%%%%%%%

\vspace{-1mm}
\section{Combining topological and ordinary finite volume effects}
\vspace{-1mm}

The results in Section 3 were obtained for volumes where ordinary
finite volume effects are negligible. This was rather easy, as SU(2) 
theory is a special case since even the lightest glueball
has a considerable mass, which implies strongly
suppressed ordinary finite volume effects. As indicated in 
Section \ref{Ordinary-finite-volume}, and applied in Section 
\ref{sec:Topological-finite-volume}, one has to discard volumes 
with $\hat L<14$, {\it i.e.}\ box sizes $L \lesssim 1~{\rm fm}$. 
On the other hand, in QCD the lightest particle is the pion,
which is much lighter than any glueball in SU(2) Yang-Mills theory. 
Therefore one expects persistent 
ordinary finite volume effects up to larger lattices.

\begin{figure}
\begin{centering}
\includegraphics[width=10cm]{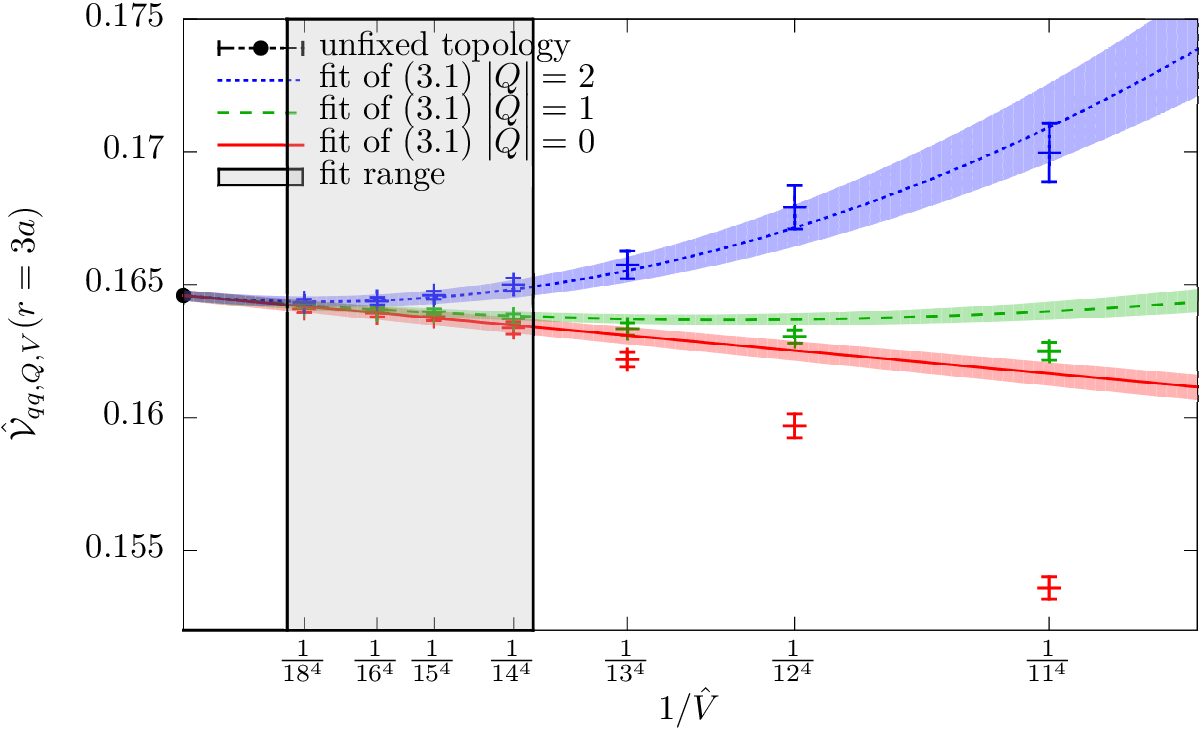}
\par\end{centering}
\protect\caption{\label{BCNW-finite-volume-effects}$\hat{\mathcal{V}}_{q 
\bar{q},Q,V}(\hat r = 3)$ as a function of $1/\hat{V}$. The curves 
represent the fit of eq.\ (\protect\ref{eq:BCNW}) to the lattice 
results in the four largest volumes, $\hat{V} = 14^4, 15^4, 16^4, 18^4$. 
There is are sizable discrepancies between these curves and the lattice 
data in the smaller volumes, $\hat{V} = 11^4, 12^4, 13^4$.}
\vspace*{-2mm}
\end{figure}

In order to understand how eq.\ (\ref{eq:BCNW}) is affected by ordinary
finite volume effects, we also computed the static $q\bar{q}$-potential
for smaller lattices, $\hat L < 14$, which are affected by significant 
finite volume effects.
In Figure \ref{BCNW-finite-volume-effects} we show results
for seven volumes in the range $11^{4} \dots 18^{4}$, 
for $|Q|=0,1$ and $2$. The lines represent the fit
of the formula (\ref{eq:BCNW}) in volumes $\hat V > 13^{4}$. 
For these volumes, this equation is well compatible with our 
data. For smaller volumes, however, the discrepancies
are significant and clearly show that ordinary finite volume
effects have to be taken into account.
%%%%%%%%%%%%%%%%%%%%%%%%%%%%%%%%%%%%%%%%%%%%%%%%%%%%%%
%%%%%%%%%%%%%%%%%%%%%%%%%%%%%%%%%%%%%%%%%%%%%%%%%%%%%%
Therefore it is highly desirable to combine eq.\ (\ref{eq:BCNW}) with
eq.\ (\ref{eq:SU(N)}) or (\ref{eq:QCD}), in order to obtain a formula
which captures both ordinary {\em and} topological finite size
effects. Comparing the topological sectors, one notices that
the difference between the curve and the data is larger for topological
charge $Q=0$ than for  $|Q|=1$, $2$.
For the latter the finite volume effects seem negligible. This different 
behaviour underlines the topological charge dependence of ordinary finite 
volume effects at fixed topology. 
%To include this effect,
%we have to study this dependence and compute at least
%the next to leading order equations (leading order corresponds
%to a simple addition of the two effects). 

To derive such expressions, one first has to compute ordinary finite 
volume effects also at non-vanishing $\theta$ angles, using 
equations analogous to (\ref{eq:SU(N)}) or (\ref{eq:QCD}). 
To this end, we employed the L\"{u}scher method \cite{Luscher:1985dn}, 
which is universal and therefore applicable to SU($N$) Yang-Mills theory. 
Starting from the finite volume equation at fixed $\theta$, one has 
to perform a calculus similar the one leading to eq.\ (\ref{eq:BCNW}),
see Ref.\ \cite{Dromard:2014ela}. 
In Yang-Mills theory, the expression, which captures both types of finite 
volume effects, is
%%%%%%%%%%%%%%%%%%%%%%%%%%%%%%%%%%%%%%%%%%%%%%%%%%%%%%
\vspace*{-2mm}
\begin{equation}
\begin{aligned}M_{Q,L} & 
\approx M+\frac{1}{2\chi_{t}V}M^{(2)}\left(1-\frac{Q^{2}}{\chi_{t}V}\right)-\frac{A}{L}
\exp \left(-\frac{\sqrt{3}m_{\rm g}L}{2}\right)\\
 & 
+\frac{1}{2\chi_{t}V}\left(B-A\sqrt{3}m^{(2)}_{\rm g}\right)
\left(1-\frac{Q^{2}}{\chi_{t}V}\right) \exp
\left(-\frac{\sqrt{3}m_{\rm g}L}{2}\right) \ ,
\end{aligned}
\label{eq5}
\vspace*{-2mm}
\end{equation}
%%%%%%%%%%%%%%%%%%%%%%%%%%%%%%%%%%%%%%%%%%%%%%%%%%%%%%
%%%%%%%%%%%%%%%%%%%%%%%%%%%%%%%%%%%%%%%%%%%%%%%%%%%%%%
where $M_{Q,L}$ is a hadron mass in a finite volume in one
topological sector, whereas $M$ and $m_{\rm g}$ are the physical 
mass (at infinite  volume and $\theta=0$) of the hadron and the 
$0^{++}$ glueball.
%$L$ is the spatial extent and $V$ the space-time volume. 
$A$ and $B$ are coefficients, which are independent of $V$ and $Q$,
and $(2)$ denotes the second derivative with respect to $\theta$ at $\theta=0$. 
The corresponding formula for the pion mass in QCD reads
%%%%%%%%%%%%%%%%%%%%%%%%%%%%%%%%%%%%%%%%%%%%%%%%%%%%%%  
\begin{equation}
\small{
\begin{aligned}m_{\pi,Q,L} & 
\approx {m_{\pi}+\frac{m_{\pi}^{(2)}}{2\chi_{t}V}\left(1-\frac{Q^{2}}{\chi_{t
}V}\right)}{+\frac{3}{16\pi^{2}}\frac{m_{\pi}^{2}}{F_{\pi}^{2}}
\frac{K_{1}(x)}{x}}+\frac{m_{\pi}^{(2)}E}{2\chi_{t}V}\left(1-\frac{2Q^{2}}{\chi_
{t}V}\right)\frac{K_{1}(x)}{x}\\ 
& 
+\frac{3}{32\pi^{2}\chi_{t}V}\frac{m_{\pi}^{2}}{F_{\pi}^{2}}\left(1-\frac{Q^{2} 
} 
{\chi_{t}V}\right)\left[\left(\frac{m_{\pi}^{(2)}}{m_{\pi}}-2\frac{F_{\pi}^{(2)}
}{F_{\pi}}\right)\frac{K_{1}(x)}{x}{-\frac{1}{2}\frac{m_{\pi}^{(2)}}{m_{\pi}}
\Big( K_{0}(x)+K_{2}(x)\Big) }\right] \ ,
\end{aligned}}
\label{eq:6}
\end{equation}
%%%%%%%%%%%%%%%%%%%%%%%%%%%%%%%%%%%%%%%%%%%%%%%%%%%%%%
%%%%%%%%%%%%%%%%%%%%%%%%%%%%%%%%%%%%%%%%%%%%%%%%%%%%%%
%In eq.\ (\ref{eq:6}), $m_{\pi,Q,L}$ is the pion mass at finite volume
%and for one topological charge, $m_{\pi}$ is physical the pion mass, 
%$V$ the volume, 
where $F_{\pi}$ is the pion decay constant, $E$ is another constant
(independent of $V$ and $Q$), $K_{n}$ is the modified Bessel function 
%of order $n$ and finally 
and $x=m_{\pi}L$. 

\vspace{-1mm}
\section{Numerical test in SU(2) gauge theory}
\vspace{-1mm}

\begin{figure}
\begin{centering}
\includegraphics[width=10cm]{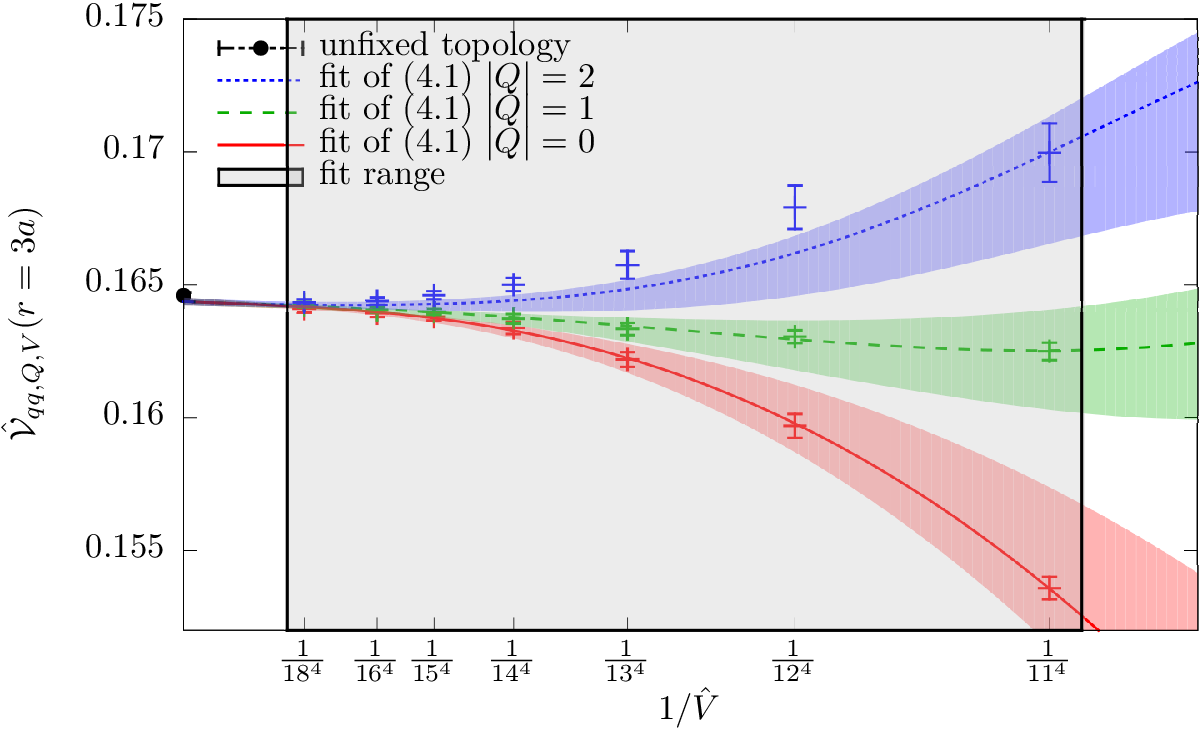}
\par\end{centering}
\protect\caption{\label{NLO FVE}$\hat{\mathcal{V}}_{q \bar{q},Q,V}(\hat r = 3)$ 
as a function of $1/\hat{V}$. The curves represent the fit of eq.\ 
(\protect\ref{eq5}) to the lattice results in all volumes 
$\hat{V} = 11^4, \dots, 18^4$. 
We obtain very good agreement, even at small volumes and $Q = 0$. }
\vspace{-2mm}
\end{figure}

Figure \ref{NLO FVE} shows the same data points as Figure 
\ref{BCNW-finite-volume-effects}, but now with a fit %performed
according to eq.\ (\ref{eq5}), which includes the ordinary finite 
volume effects. The fitting parameters are
$\hat M, \hat M^{(2)},\hat m_{\rm g}, \hat m^{(2)}_{\rm g},
A,B \text{ and }\hat \chi_{t}$.
We see that eq.\ (\ref{eq5}) matches the data very well, with
$\chi^{2}/{\rm d.o.f.}<1$; the explicit
results are given in Table \ref{Results-NLO}.
They all agree with the literature, albeit with considerable
errors for $\hat m_{\rm g}$ and $\hat \chi_{t}$. In particular, 
for the topological susceptibility the error is larger than 20\%, 
hence for that quantity a different approach should be used, such 
as the method discussed in Section \ref{sec:Topological-finite-volume}.
Nevertheless the evaluation of $\hat{\mathcal{V}}_{q\bar{q}}$
is achieved to a high precision, with an error below $0.1\%$. 
%%%%%%%%%%%%%%%%%%%%%%%%%%%%%%%%%%%%%%%%%%%%%%%%%%%%%%
%%%%%%%%%%%%%%%%%%%%%%%%%%%%%%%%%%%%%%%%%%%%%%%%%%%%%%
%%%%%%%%%%%%%%%%%%%%%%%%%%%%%%%%%%%%%%%%%%%%%%%%%%%%%%
\begin{table}
\begin{centering}
\begin{tabular}{|c|c|c|c|}
\hline 
 & $\hat{\mathcal{V}}_{qq}(\hat r=3)$ & $\hat m_{\rm g}$ 
 & $\hat \chi_{t} \times10^{5} $\tabularnewline
\hline 
\hline 
fit based on eq.\ (\ref{eq5}) 
& $0.16437(15)$ & $0.67(10)$ & $9.5(2.0)$\tabularnewline
\hline 
result at unfixed topology \cite{Teper:1998kw,deForcrand:1997sq} & $0.16455(7)$ & 
$0.723(23)$ & $7.0(0.9)$\tabularnewline
\hline 
\end{tabular}
\par\end{centering}
\protect\caption{\label{Results-NLO}Results for the static potential 
$\hat{\mathcal{V}}_{q \bar{q}}(\hat r = 3)$, the mass $\hat{m}_{\rm g}$ of 
the $J^{PC} = 0^{++}$ glueball and the topological susceptibility
$\hat{\chi_{t}}$, obtained by a fit of eq.\ (\protect\ref{eq5}) to fixed
topology lattice results for $\hat{\mathcal{V}}_{q \bar{q},Q,V}(\hat r = 3)$.}
%\vspace*{-2mm}
\end{table}
%%%%%%%%%%%%%%%%%%%%%%%%%%%%%%%%%%%%%%%%%%%%%%%%%%%%%% 

To conclude, %in order to apply this method for QCD simulations, 
we have included ordinary finite volume effects in a relation between a 
hadron mass at fixed topology and finite volume, and the corresponding physical mass. We
successfully tested this new relation in SU(2) Yang-Mills theory for
the static quark-antiquark potential. Due to this extension, the method
becomes more promising in QCD applications. This is currently under 
investigation.

\vspace{-1mm}
\section*{Acknowledgements}
\vspace{-1mm}

%{\bf Acknowledgements} \ \
 A.D. and M.W. acknowledge support by the Emmy Noether Programme of the DFG 
(German Research Foundation), grant WA 3000/1-1.
W.B., U.G.\ and H.M.-D.\ acknowledge support by the Consejo Nacional 
de Ciencia y Tecnolog\'{\i}a (CONACYT) through projects CB-2010/155905
and CB-2013/222812, as well as DGAPA-UNAM, grant IN107915.
 
This work was supported in part by the Helmholtz International Center for FAIR 
within the framework of the LOEWE program launched by the State of Hesse. 
Calculations on the LOEWE-CSC high-performance computer of Johann Wolfgang 
Goethe-University Frankfurt am Main were conducted for this research. We would 
like to thank HPC-Hessen, funded by the State Ministry of Higher Education, 
Research and the Arts, for programming advice.


\begin{thebibliography}{99}

\bibitem{Luscher:2011kk}
  M.~L\"uscher and S.~Schaefer,
  %``Lattice QCD without topology barriers,''
  JHEP {\bf 1107}, 036 (2011)
  [arXiv:1105.4749 [hep-lat]].
  %%CITATION = ARXIV:1105.4749;%%

\bibitem{Aoki:2008tq}
  S.~Aoki {\it et al.} [JLQCD Collaboration],
  %``Two-flavor QCD simulation with exact chiral symmetry,''
  Phys.\ Rev.\ D {\bf 78}, 014508 (2008)
  [arXiv:0803.3197 [hep-lat]].
  %%CITATION = ARXIV:0803.3197;%%

\bibitem{Borsanyi} Sz.\ Borsanyi, % {\it et al.}, 
Z.\ Fodor, S.D.\ Katz, S.\ Krieg, T.\ Lippert, D.\ Nogradi, 
F.\ Pittler, K.K.\ Szabo and B.C. Toth,
% QCD thermodynamics with continuum extrapolated dynamical overlap fermions
arXiv:1510.03376 [hep-lat]. 

\bibitem{Cichy:2010ta}
  K.~Cichy, G.~Herdoiza and K.~Jansen,
  %``Continuum Limit of Overlap Valence Quarks on a Twisted Mass Sea,''
  Nucl.\ Phys.\ B {\bf 847}, 179 (2011)
  [arXiv:1012.4412 [hep-lat]].
  %%CITATION = ARXIV:1012.4412;%%

\bibitem{Brower:2003yx}
  R.~Brower, S.~Chandrasekharan, J.~W.~Negele and U.-J.~Wiese,
  %``QCD at fixed topology,''
  Phys.\ Lett.\ B {\bf 560}, 64 (2003)
  [hep-lat/0302005].
  %%CITATION = HEP-LAT/0302005;%%

\bibitem{Aoki:2007ka}
  S.~Aoki, H.~Fukaya, S.~Hashimoto and T.~Onogi,
  %``Finite volume QCD at fixed topological charge,''
  Phys.\ Rev.\ D {\bf 76}, 054508 (2007)
  [arXiv:0707.0396 [hep-lat]].
  %%CITATION = ARXIV:0707.0396;%%
  
  \bibitem{Dromard:2014ela}
  A.~Dromard and M.~Wagner,
  %``Extracting hadron masses from fixed topology simulations,''
  Phys.\ Rev.\ D {\bf 90}, 074505 (2014)
  [arXiv:1404.0247 [hep-lat]].
  %%CITATION = ARXIV:1404.0247;%%
  
%\cite{Bietenholz:2011ey}
\bibitem{Bietenholz:2011ey} 
  W.~Bietenholz, I.~Hip, S.~Shcheredin and J.~Volkholz,
  %``A Numerical Study of the 2-Flavour Schwinger Model with Dynamical Overlap 
%Hypercube Fermions,''
  Eur.\ Phys.\ J.\ C {\bf 72}, 1938 (2012)
  [arXiv:1109.2649 [hep-lat]].
  %%CITATION = ARXIV:1109.2649;%%
  %16 citations counted in INSPIRE as of 24 Apr 2015

\bibitem{Bietenholz:2008rj}
  W.~Bietenholz and I.~Hip,
  %``Topological Summation of Observables Measured with Dynamical Overlap 
%Fermions,''
  PoS LATTICE {\bf 2008}, 079 (2008)
  [arXiv:0808.3049 [hep-lat]].
  %%CITATION = ARXIV:0808.3049;%%

\bibitem{Bietenholz:2012sh}
  W.~Bietenholz and I.~Hip,
  %``Topological Summation in Lattice Gauge Theory,''
  J.\ Phys.\ Conf.\ Ser.\ {\bf 378}, 012041 (2012)
  [arXiv:1201.6335 [hep-lat]].
  %%CITATION = ARXIV:1201.6335;%%

\bibitem{Dromard:2013wja}
  A.~Dromard and M.~Wagner,
  %``Studying and removing effects of fixed topology in a quantum mechanical 
%model,''
  PoS LATTICE {\bf 2013}, 339 (2014)
  [arXiv:1309.2483 [hep-lat]].
  %%CITATION = ARXIV:1309.2483;%%

\bibitem{Czaban:2013haa}
  C.~Czaban and M.~Wagner,
  %``Lattice study of the Schwinger model at fixed topology,''
  PoS LATTICE {\bf 2013}, 465 (2013)
  [arXiv:1310.5258 [hep-lat]].
  %%CITATION = ARXIV:1310.5258;%%
  
  %\cite{Bautista:2014tba}
\bibitem{Bautista:2014tba} 
  I.~Bautista {\it et al.},
  %``Interpretation of topologically restricted measurements in lattice 
%sigma-models,''
  arXiv:1402.2668 [hep-lat].
  %%CITATION = ARXIV:1402.2668;%%
  %5 citations counted in INSPIRE as of 24 Apr 2015

\bibitem{Czaban:2014gva}
  C.~Czaban, A.~Dromard and M.~Wagner,
  %``Studying and removing effects of fixed topology,''
  Acta Phys.\ Polon.\ Supp.\ {\bf 7}, 551 (2014)
  [arXiv:1404.3597 [hep-lat]].
  %%CITATION = ARXIV:1404.3597;%%

\bibitem{Dromard:2014gma}
  A.~Dromard, C.~Czaban and M.~Wagner,
  %``Hadron masses from fixed topology simulations: parity partners and SU(2) 
%Yang-Mills results,''
  PoS LATTICE {\bf 2014}, 321 (2014)
  [arXiv:1410.4333 [hep-lat]].
  
  %\cite{Gerber:2014bia}
\bibitem{Gerber:2014bia} 
  U.~Gerber {\it et al.},
  %``Extracting Physics from Topologically Frozen Markov Chains,''
  PoS LATTICE {\bf 2014} (2014)
  [arXiv:1410.0426 [hep-lat]].
  %%CITATION = ARXIV:1410.0426;%%
  %2 citations counted in INSPIRE as of 24 Apr 2015

\bibitem{Bautista:2015yza}
  I.~Bautista {\it et al.},
  %``Measuring the Topological Susceptibility in a Fixed Sector: Results for 
%Sigma Models,''
  arXiv:1503.06853 [hep-lat].
  
  %\cite{Dromard:2015oqa}
\bibitem{Dromard:2015oqa} 
  A.~Dromard, W.~Bietenholz, U.~Gerber, H.~Mej\'{i}a-D\'{i}az and M.~Wagner,
  %``Simulations at fixed topology: fixed topology versus ordinary finite volume 
%corrections,''
  Acta Phys.\ Polon.\ Supp.\  {\bf 8}, no. 2, 391 (2015)
  [arXiv:1505.03435 [hep-lat]].
  %%CITATION = ARXIV:1505.03435;%%

\bibitem{Luscher:1985dn}
  M.~L\"uscher,
  %``Volume Dependence of the Energy Spectrum in Massive Quantum Field Theories. 
%1. Stable Particle States,''
  Commun.\ Math.\ Phys.\ {\bf 104}, 177 (1986).
  %%CITATION = CMPHA,104,177;%%

\bibitem{Teper:1998kw} 
  M.~J.~Teper,
  %``Glueball masses and other physical properties of SU(N) gauge theories in D= 
%(3+1): A Review of lattice results for theorists,''
  hep-th/9812187.
  %%CITATION = HEP-TH/9812187;%%

\bibitem{deForcrand:1997sq}
  P.~de Forcrand, M.~Garc\'{i}a P\'{e}rez and I.~O.~Stamatescu,
  %``Topology of the SU(2) vacuum: A Lattice study using improved cooling,''
  Nucl.\ Phys.\ B {\bf 499}, 409 (1997)
  [hep-lat/9701012].
  %%CITATION = HEP-LAT/9701012;%%

\end{thebibliography}
\end{document}